\RequirePackage{lineno}

  \documentclass[aps,prd,twocolumn,superscriptaddress,showpacs]{revtex4}

\usepackage{graphicx}
\usepackage{dcolumn}
\usepackage{bm}
\usepackage{color}

\newcommand{\ppbar}{$\bar{p}p$}                                                                                             %

\newcommand{\dedx}{\ensuremath{\mathit{dE/dx}}}                                 
\newcommand{\lumifb}{ fb$^{-1}$}

\newcommand{\sigLxy}{\ensuremath{L_{xy}(B)/\sigma_{L_{xy}(B)}}}
\newcommand{\LxyD}{\ensuremath{L_{xy}(D)}}

\begin{document}

 \title{Measurements of branching fraction ratios and $\mathit{CP}$ asymmetries in $B^{\pm} \rightarrow D_{\mathit{CP}}K^{\pm}$ decays in hadron collisions}
 \affiliation{Institute of Physics, Academia Sinica, Taipei, Taiwan 11529, Republic of China} 
\affiliation{Argonne National Laboratory, Argonne, Illinois 60439} 
\affiliation{University of Athens, 157 71 Athens, Greece} 
\affiliation{Institut de Fisica d'Altes Energies, Universitat Autonoma de Barcelona, E-08193, Bellaterra (Barcelona), Spain} 
\affiliation{Baylor University, Waco, Texas  76798} 
\affiliation{Istituto Nazionale di Fisica Nucleare Bologna, $^{cc}$University of Bologna, I-40127 Bologna, Italy} 
\affiliation{Brandeis University, Waltham, Massachusetts 02254} 
\affiliation{University of California, Davis, Davis, California  95616} 
\affiliation{University of California, Los Angeles, Los Angeles, California  90024} 
\affiliation{University of California, San Diego, La Jolla, California  92093} 
\affiliation{University of California, Santa Barbara, Santa Barbara, California 93106} 
\affiliation{Instituto de Fisica de Cantabria, CSIC-University of Cantabria, 39005 Santander, Spain} 
\affiliation{Carnegie Mellon University, Pittsburgh, PA  15213} 
\affiliation{Enrico Fermi Institute, University of Chicago, Chicago, Illinois 60637}
\affiliation{Comenius University, 842 48 Bratislava, Slovakia; Institute of Experimental Physics, 040 01 Kosice, Slovakia} 
\affiliation{Joint Institute for Nuclear Research, RU-141980 Dubna, Russia} 
\affiliation{Duke University, Durham, North Carolina  27708} 
\affiliation{Fermi National Accelerator Laboratory, Batavia, Illinois 60510} 
\affiliation{University of Florida, Gainesville, Florida  32611} 
\affiliation{Laboratori Nazionali di Frascati, Istituto Nazionale di Fisica Nucleare, I-00044 Frascati, Italy} 
\affiliation{University of Geneva, CH-1211 Geneva 4, Switzerland} 
\affiliation{Glasgow University, Glasgow G12 8QQ, United Kingdom} 
\affiliation{Harvard University, Cambridge, Massachusetts 02138} 
\affiliation{Division of High Energy Physics, Department of Physics, University of Helsinki and Helsinki Institute of Physics, FIN-00014, Helsinki, Finland} 
\affiliation{University of Illinois, Urbana, Illinois 61801} 
\affiliation{The Johns Hopkins University, Baltimore, Maryland 21218} 
\affiliation{Institut f\"{u}r Experimentelle Kernphysik, Karlsruhe Institute of Technology, D-76131 Karlsruhe, Germany} 
\affiliation{Center for High Energy Physics: Kyungpook National University, Daegu 702-701, Korea; Seoul National University, Seoul 151-742, Korea; Sungkyunkwan University, Suwon 440-746, Korea; Korea Institute of Science and Technology Information, Daejeon 305-806, Korea; Chonnam National University, Gwangju 500-757, Korea; Chonbuk National University, Jeonju 561-756, Korea} 
\affiliation{Ernest Orlando Lawrence Berkeley National Laboratory, Berkeley, California 94720} 
\affiliation{University of Liverpool, Liverpool L69 7ZE, United Kingdom} 
\affiliation{University College London, London WC1E 6BT, United Kingdom} 
\affiliation{Centro de Investigaciones Energeticas Medioambientales y Tecnologicas, E-28040 Madrid, Spain} 
\affiliation{Massachusetts Institute of Technology, Cambridge, Massachusetts  02139} 
\affiliation{Institute of Particle Physics: McGill University, Montr\'{e}al, Qu\'{e}bec, Canada H3A~2T8; Simon Fraser University, Burnaby, British Columbia, Canada V5A~1S6; University of Toronto, Toronto, Ontario, Canada M5S~1A7; and TRIUMF, Vancouver, British Columbia, Canada V6T~2A3} 
\affiliation{University of Michigan, Ann Arbor, Michigan 48109} 
\affiliation{Michigan State University, East Lansing, Michigan  48824}
\affiliation{Institution for Theoretical and Experimental Physics, ITEP, Moscow 117259, Russia} 
\affiliation{University of New Mexico, Albuquerque, New Mexico 87131} 
\affiliation{Northwestern University, Evanston, Illinois  60208} 
\affiliation{The Ohio State University, Columbus, Ohio  43210} 
\affiliation{Okayama University, Okayama 700-8530, Japan} 
\affiliation{Osaka City University, Osaka 588, Japan} 
\affiliation{University of Oxford, Oxford OX1 3RH, United Kingdom} 
\affiliation{Istituto Nazionale di Fisica Nucleare, Sezione di Padova-Trento, $^{dd}$University of Padova, I-35131 Padova, Italy} 
\affiliation{LPNHE, Universite Pierre et Marie Curie/IN2P3-CNRS, UMR7585, Paris, F-75252 France} 
\affiliation{University of Pennsylvania, Philadelphia, Pennsylvania 19104}
\affiliation{Istituto Nazionale di Fisica Nucleare Pisa, $^{ee}$University of Pisa, $^{ff}$University of Siena and $^{gg}$Scuola Normale Superiore, I-56127 Pisa, Italy} 
\affiliation{University of Pittsburgh, Pittsburgh, Pennsylvania 15260} 
\affiliation{Purdue University, West Lafayette, Indiana 47907} 
\affiliation{University of Rochester, Rochester, New York 14627} 
\affiliation{The Rockefeller University, New York, New York 10021} 
\affiliation{Istituto Nazionale di Fisica Nucleare, Sezione di Roma 1, $^{hh}$Sapienza Universit\`{a} di Roma, I-00185 Roma, Italy} 

\affiliation{Rutgers University, Piscataway, New Jersey 08855} 
\affiliation{Texas A\&M University, College Station, Texas 77843} 
\affiliation{Istituto Nazionale di Fisica Nucleare Trieste/Udine, I-34100 Trieste, $^{ii}$University of Trieste/Udine, I-33100 Udine, Italy} 
\affiliation{University of Tsukuba, Tsukuba, Ibaraki 305, Japan} 
\affiliation{Tufts University, Medford, Massachusetts 02155} 
\affiliation{Waseda University, Tokyo 169, Japan} 
\affiliation{Wayne State University, Detroit, Michigan  48201} 
\affiliation{University of Wisconsin, Madison, Wisconsin 53706} 
\affiliation{Yale University, New Haven, Connecticut 06520} 
\author{T.~Aaltonen}
\affiliation{Division of High Energy Physics, Department of Physics, University of Helsinki and Helsinki Institute of Physics, FIN-00014, Helsinki, Finland}
\author{J.~Adelman}
\affiliation{Enrico Fermi Institute, University of Chicago, Chicago, Illinois 60637}
\author{B.~\'{A}lvarez~Gonz\'{a}lez$^v$}
\affiliation{Instituto de Fisica de Cantabria, CSIC-University of Cantabria, 39005 Santander, Spain}
\author{S.~Amerio$^{dd}$}
\affiliation{Istituto Nazionale di Fisica Nucleare, Sezione di Padova-Trento, $^{dd}$University of Padova, I-35131 Padova, Italy} 

\author{D.~Amidei}
\affiliation{University of Michigan, Ann Arbor, Michigan 48109}
\author{A.~Anastassov}
\affiliation{Northwestern University, Evanston, Illinois  60208}
\author{A.~Annovi}
\affiliation{Laboratori Nazionali di Frascati, Istituto Nazionale di Fisica Nucleare, I-00044 Frascati, Italy}
\author{J.~Antos}
\affiliation{Comenius University, 842 48 Bratislava, Slovakia; Institute of Experimental Physics, 040 01 Kosice, Slovakia}
\author{G.~Apollinari}
\affiliation{Fermi National Accelerator Laboratory, Batavia, Illinois 60510}
\author{A.~Apresyan}
\affiliation{Purdue University, West Lafayette, Indiana 47907}
% ...

\collaboration{CDF Collaboration\footnote{With visitors from $^a$University of Massachusetts Amherst, Amherst, Massachusetts 01003,
$^b$Universiteit Antwerpen, B-2610 Antwerp, Belgium, 
$^c$University of Bristol, Bristol BS8 1TL, United Kingdom,
$^d$Chinese Academy of Sciences, Beijing 100864, China, 
$^e$Istituto Nazionale di Fisica Nucleare, Sezione di Cagliari, 09042 Monserrato (Cagliari), Italy,
$^f$University of California Irvine, Irvine, CA  92697, 
$^g$University of California Santa Cruz, Santa Cruz, CA  95064, 
$^h$Cornell University, Ithaca, NY  14853, 
$^i$University of Cyprus, Nicosia CY-1678, Cyprus, 
$^j$University College Dublin, Dublin 4, Ireland,
$^k$University of Edinburgh, Edinburgh EH9 3JZ, United Kingdom, 
$^l$University of Fukui, Fukui City, Fukui Prefecture, Japan 910-0017
$^m$Kinki University, Higashi-Osaka City, Japan 577-8502
$^n$Universidad Iberoamericana, Mexico D.F., Mexico,
$^o$University of Iowa, Iowa City, IA  52242,
$^p$Kansas State University, Manhattan, KS 66506
$^q$Queen Mary, University of London, London, E1 4NS, England,
$^r$University of Manchester, Manchester M13 9PL, England,
$^s$Muons, Inc., Batavia, IL 60510, 
$^t$Nagasaki Institute of Applied Science, Nagasaki, Japan, 
$^u$University of Notre Dame, Notre Dame, IN 46556,
$^v$University de Oviedo, E-33007 Oviedo, Spain, 
$^w$Texas Tech University, Lubbock, TX  79609, 
$^x$IFIC(CSIC-Universitat de Valencia), 56071 Valencia, Spain,
$^y$Universidad Tecnica Federico Santa Maria, 110v Valparaiso, Chile,
$^z$University of Virginia, Charlottesville, VA  22906
$^{aa}$Bergische Universit\"at Wuppertal, 42097 Wuppertal, Germany,
$^{bb}$Yarmouk University, Irbid 211-63, Jordan
$^{jj}$On leave from J.~Stefan Institute, Ljubljana, Slovenia, 
}}
\noaffiliation

 
  \date{October 28, 2009}
  
  \begin{abstract}
  We reconstruct $B^{\pm} \rightarrow D K^{\pm}$ decays in a data
  sample collected by the CDF II detector at the Tevatron collider
  corresponding to 1 fb$^{-1}$ of integrated luminosity.  We
  select decay modes where the $D$ meson decays to either $K^- \pi^+$
  (flavor eigenstate) or $K^- K^+, \pi^- \pi^+$ ($\mathit{CP}$-even
  eigenstates), and measure the direct $\mathit{CP}$ asymmetry $A_{\mathit{CP+}} =
  0.39\pm 0.17(\rm{stat})\pm 0.04(\rm{syst})$, and the double ratio of
  $\mathit{CP}$-even to flavor eigenstate branching fractions $R_{\mathit{CP+}} =
  1.30\pm 0.24(\rm{stat})\pm 0.12(\rm{syst})$.  These measurements will
  improve the determination of the Cabibbo-Kobayashi-Maskawa angle $\gamma$.
  They are performed here for the first time using data from hadron
  collisions.
 \end{abstract}
  
  
  
  
  \maketitle
  
  
  
  The measurement of $\mathit{CP}$ asymmetries and branching ratios of $B^{-} \rightarrow D
  K^{-}$~\cite{chargedecay} decay modes allows a theoretically-clean
  extraction of the CKM angle $\gamma = \arg(-V_{ud}V^*_{ub}/V_{cd}V^*_{cb})$, a fundamental parameter of the standard model~\cite{Kobayashi}.
  In these decays the interference between the tree amplitudes of the 
   $b\to c\bar{u}s$ and $b\to u\bar{c}s$ processes leads to observables that depend on their relative weak 
  phase ($\gamma$), their relative strong phase ($\delta_{B}$), and 
  the magnitude ratio $r_{B}=\left| \frac{A(b\to u)}{A(b\to c)} 
  \right| $. These quantities can all be extracted from data by combining several experimental observables. This can be achieved in
  several ways, from a variety of $D$ decay
  channels~\cite{Gronau, Atwood, Giri}. 
  
  An accurate knowledge of the value of $\gamma$ is instrumental in establishing  the possible presence of additional non-standard model $\mathit{CP}$-violating phases in higher-order diagrams~\cite{fleischer,dunietz}.
 Its current determination is based on a combination of several $B \rightarrow D
  K$ measurements performed in $e^+e^-$ collisions at the $\Upsilon$(4S) resonance~\cite{hfag,babar,belle} and its uncertainty is between $12$ and  $30$ deg, depending on the method~\cite{ckmutfit}. This 
  uncertainty is almost completely determined by the limited size of the data 
  samples available, with theoretical uncertainties playing a negligible role ($\sim 1\%$). 
  The large production of $B$ mesons available at hadron
  colliders could offer a unique opportunity to improve 
  the current experimental determination of the angle $\gamma$. 
  However, the feasibility of this kind of measurement in the larger background conditions of hadronic collisions has never been demonstrated.
  
  
  In this paper we describe the first measurement of the 
  branching fraction ratios and $\mathit{CP}$ asymmetries
  of $B^{-} \rightarrow D K^{-}$ modes
  performed in hadron collisions, based on an integrated 
  luminosity of 1\lumifb\ of 
  \ppbar\ collisions at $\sqrt{s}$ = 1.96 TeV collected 
  by the upgraded Collider Detector (CDF II) at the Fermilab 
  Tevatron. 
  We reconstruct events where the $D$ meson decays to the flavor-specific 
  mode $K^- \pi^+$ ($D^0_{f}$), or to one of the $\mathit{CP}$-even modes
  $K^- K^+$ and $\pi^- \pi^+$ [$D_{\mathit{CP+}} = (D^0+\overline{D}^0)/\sqrt{2}$]. 
  From these modes, the following observables can be defined:
      \begin{equation}\label{acp_eq}
    A_{\mathit{CP+}} = \frac{\mathcal{B}(B^- \rightarrow D_{\mathit{CP+}} K^-)-\mathcal{B}(B^+
    \rightarrow D_{\mathit{CP+}} K^+)}{\mathcal{B}(B^- \rightarrow D_{\mathit{CP+}}
    K^-)+\mathcal{B}(B^+ \rightarrow D_{\mathit{CP+}} K^+)},
    \end{equation}
    \begin{equation}\label{rcp_eq}
    R_{\mathit{CP+}} = 2 \frac{\mathcal{B}(B^- \rightarrow D_{\mathit{CP+}} K^-)+\mathcal{B}(B^+ \rightarrow
    D_{\mathit{CP+}} K^+)}{\mathcal{B}(B^-
 \rightarrow D^0_{f} K^-)+\mathcal{B}(B^+ \rightarrow
 \overline{D}^0_{f} K^+)}.
  \end{equation}
  
 With the assumption of no $\mathit{CP}$ violation in $D^0$ decays, and neglecting 
 $D^0$$-$$\overline{D}^0$ mixing \cite{d0mixing},
 these quantities are related to the CKM angle $\gamma$ by the
 equations~\cite{Gronau}
    \begin{eqnarray}
    R_{\mathit{CP+}}=1+r_B^2+2r \cos{\delta_B}\cos{\gamma},\\
    A_{\mathit{CP+}}=2r_B \sin{\delta_B \sin{\gamma}/R_{\mathit{CP+}}}.
  \end{eqnarray} 
  
For our measurements we adopt the usual approximation $R_{\mathit{CP+}} \sim \frac{R_{+}}{R}$, which is valid up to a term $r \cdot |V_{us}V_{cd}/V_{ud}V_{cs}| \simeq 0.01$~\cite{Gronau_new},
 where {\small \begin{eqnarray} R &=& \frac{\mathcal{B}(B^- \rightarrow
 D^0_{f} K^-)+\mathcal{B}(B^+ \rightarrow \overline{D}^0_{f} K^+)}{\mathcal{B}(B^-
 \rightarrow D^0_{f} \pi^-)+\mathcal{B}(B^+ \rightarrow
 \overline{D}^0_{f} \pi^+)},\\
    R_+ &=& \frac{\mathcal{B}(B^- \rightarrow D_{\mathit{CP+}} K^-)+\mathcal{B}(B^+ \rightarrow
    D_{\mathit{CP+}} K^+)}{\mathcal{B}(B^- \rightarrow D_{\mathit{CP+}} \pi^-)+\mathcal{B}(B^+
    \rightarrow D_{\mathit{CP+}} \pi^+)}.
  \end{eqnarray}}
  
  The CDF II detector is a multipurpose magnetic spectrometer surrounded by
  calorimeters and muon detectors. The components relevant
  for this analysis are briefly described here. A more detailed 
  description can be found elsewhere~\cite{Acosta:2004yw}.
  Silicon microstrip detectors (SVX II and ISL)~\cite{Sill:2000zz} 
  and a cylindrical drift chamber
  (COT)~\cite{Affolder:2003ep} immersed in a 1.4 T solenoidal magnetic field
  allow reconstruction of charged particles in the pseudorapidity range 
  $\mid\eta\mid < 1.0$~\cite{CDF-coordinates}. 
  The SVX II detector consists of microstrip sensors arranged in five concentric layers with radii between 2.5 and 10.6 cm, divided into three contiguous sections along the beam direction $z$, for a total length of 90 cm. The two additional silicon layers of the ISL help to link tracks in the COT to hits in the SVX II.
   The COT has 96 measurement layers between 40 and 137 cm in radius, 
  organized into 
  alternating axial and $\pm 2^{\circ}$ stereo superlayers, and provides a resolution on the transverse momentum of charged particles $\sigma_{p_{T}}/p_{T} \simeq 
  0.15\%\, p_{T}$/(GeV/$c$). 
  The specific energy loss by ionization (\dedx) of charged particles in the COT can
  be measured from the collected charge,
  which is encoded in the output pulse width of each sense wire.
  
 
  Candidate events for this analysis are selected by a three-level
  trigger system.  At level 1, charged particles are reconstructed in the
  COT axial superlayers by a hardware processor, the extremely fast
  tracker (XFT)~\cite{Thomson:2002xp}. Two
  oppositely charged particles are required, with transverse momenta $p_T \geq 2$
  GeV/$c$ and scalar sum $p_{T1}+p_{T2}\geq 5.5$ GeV/$c$.  At level 2,
  the silicon vertex trigger (SVT) \cite{Ashmanskas:2003gf} associates
  SVX II $r-\phi$ position measurements with XFT tracks.  This provides a precise 
  measurement of the track impact parameter, $d_0$,
  which is defined as the distance of closest
  approach to the beam line.  The resolution
  of the impact parameter measurement is 50 $\mu$m for particles with 
  $p_T$ of about 2 GeV/$c$, including a $\approx 30$~$\mu$m 
  contribution due to the transverse beam size, and improves for 
  higher transverse momenta. We select $B$ hadron candidates by requiring two SVT tracks with 120
  $\leq d_0 \leq$ 1000 $\mu$m.  To reduce background from light-quark
  jet pairs, the two trigger tracks are required to have an opening
  angle in the transverse plane $2^{\circ} \leq \Delta\phi \leq
  90^{\circ}$, and to satisfy the requirement $L_{xy} > 200$ $\mu$m,
  where $L_{xy}$ is defined as the distance in the transverse plane
  from the beam line to the two-track vertex, projected onto the two-track momentum vector. The level 1 and 2 trigger requirements are then confirmed at trigger level 3, where the event is fully reconstructed.
    
  
  Reconstruction of $B^{-}$ hadrons begins by looking for a track pair that 
  is compatible with a $D^{0}$ decay. The invariant mass ($M_{D}$) of the pair is 
  required to be close to the nominal $D^{0}$ mass ($1.8 < M_{D} < 1.92$ GeV/$c^2$).
  This is checked separately for each of the four possible mass assignments 
  to the two outgoing particles: $K^{+}\pi^{-}$, $K^{-}\pi^{+}$, 
  $K^{+}K^{-}$ and $\pi^{+}\pi^{-}$. The $D^0$ candidate is combined with a negative charged track in the event with $p_T > 0.4$ GeV to form $B^{-}$ candidates. A kinematic fit of the decay is performed by constraining the two tracks forming the $D$ candidate to a common vertex and to the nominal $D^{0}$ mass,
  the $D$ candidate and the remaining track to a separate
  vertex, and the reconstructed momentum of the $B^{-}$ candidate to point back 
  to the luminous region in the transverse plane.
  
  To complete the selection, further requirements are applied on additional 
  observables: the impact parameter ($d_{B}$) of the reconstructed $B$ 
  candidate relative to the beamline; the isolation of the $B$ candidate 
  ($I_{B}$)~\cite{Isolation}; the goodness of fit of the decay vertex ($\chi^{2}_B$); 
  the transverse distance of the $D$, both relative to the beam [$L_{xy}(D)$] and to the $B$ vertex 
  [$L_{xyB}(D)$], and the significance of the $B$ hadron decay length [$\sigLxy$].
  We chose the requirement $L_{xyB}(D) > 100$ $\mu$m to reduce contamination from (nonresonant) three--body decays of the type $B^{+}\to h^{+}h^{-}h^{+}$ (from here on, we will use $h$ to indicate either $K$ or $\pi$), in which all tracks come from a common decay vertex.  
  In addition,  we reject all candidates comprising a pair of tracks with an invariant mass compatible with a $J/\psi\to\mu^+\mu^-$ decay within 2$\sigma$. 
  The threshold values for all other requirements, whose purpose is to reduce combinatorial background, were determined by an unbiased optimization procedure aimed at achieving the best resolution on $A_{\mathit{CP+}}$. This resolution was parametrized as a function of the 
  expected signal yield $S$ and background level $B$, by 
  performing repeated fits on samples of simulated data extracted 
  from the same multidimensional distribution used as likelihood function in the fit [Eq. \ref{eq:likelihood}].
  For each choice of thresholds, 
  the signal $S$ was determined by rescaling the number of observed $B^{-} \rightarrow 
  D^0_{f} \pi^{-}$, and the background $B$ was determined
  from the upper mass sidebands of each data sample ($5.4 < M_B < 5.8$ GeV/$c^2$). 
  Based on this optimization procedure, we adopted the following set of requirements: $I_B>0.65$, $\chi^2_B<13$, 
  $d_{B}< 70$~$\mu$m, $\sigLxy > 12$, and $\LxyD > 400$~$\mu$m. 

\begin{figure*}[htbp]
\begin{center}
\includegraphics[width=2in]{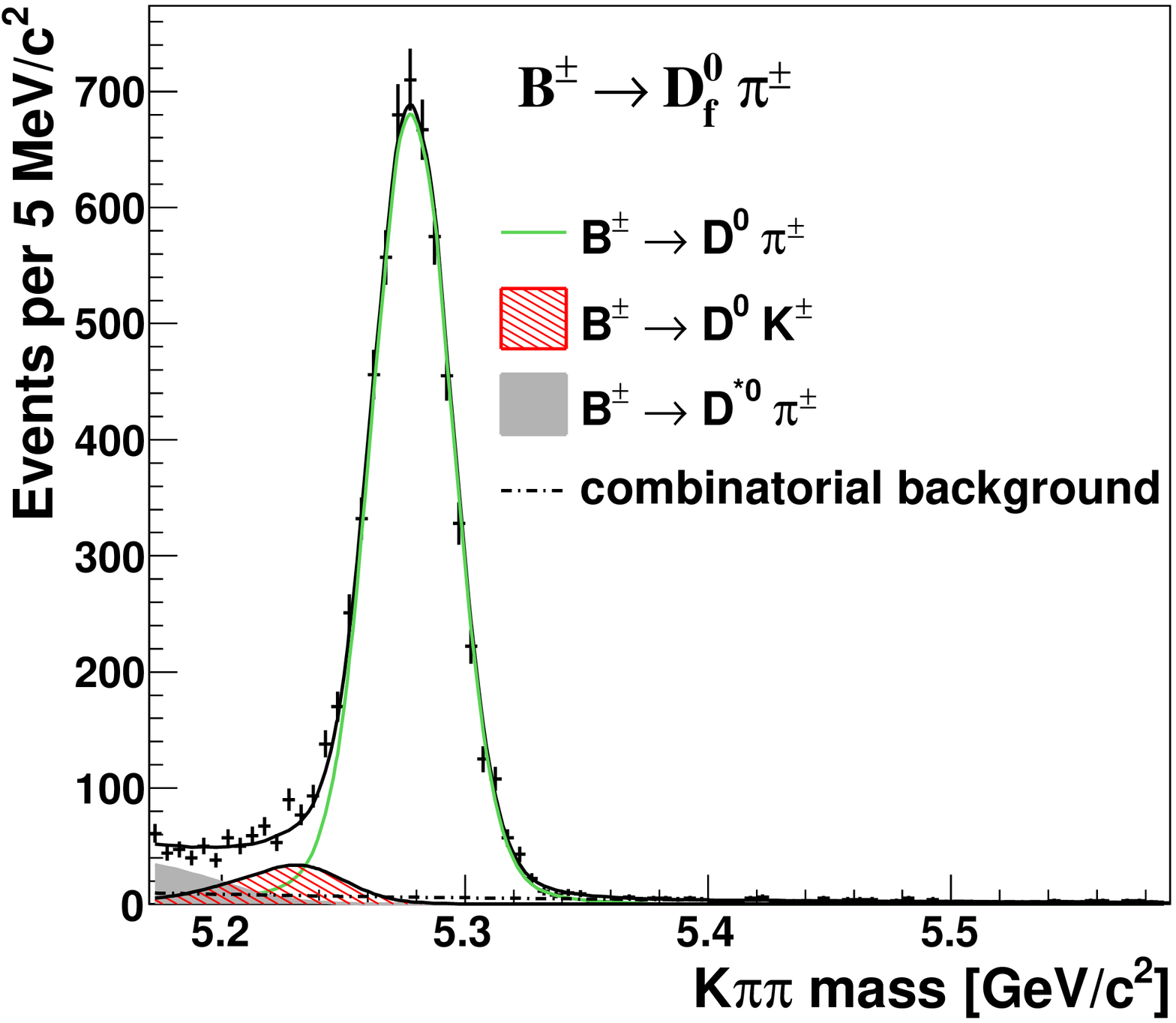}
\includegraphics[width=2in]{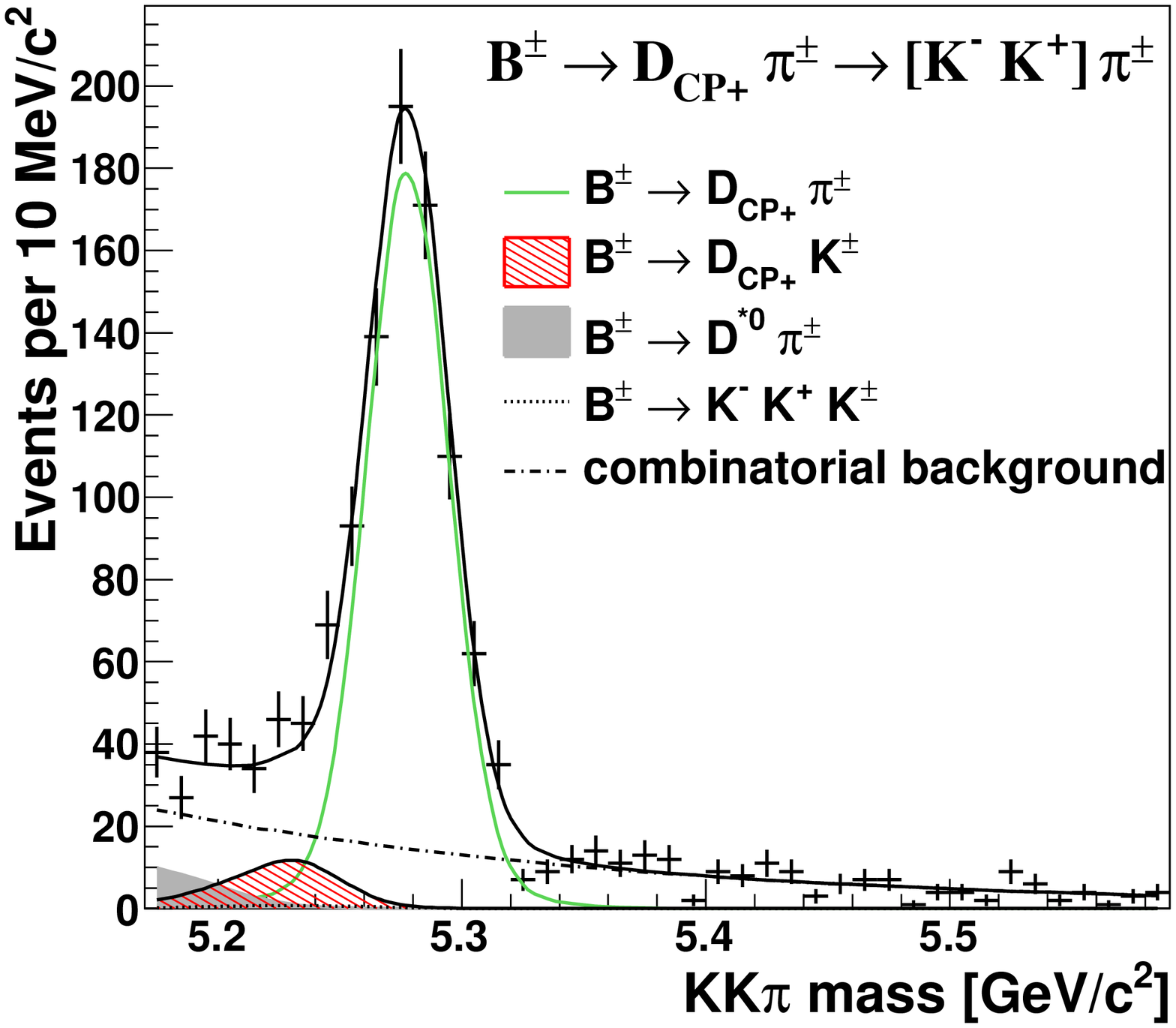}
\includegraphics[width=2in]{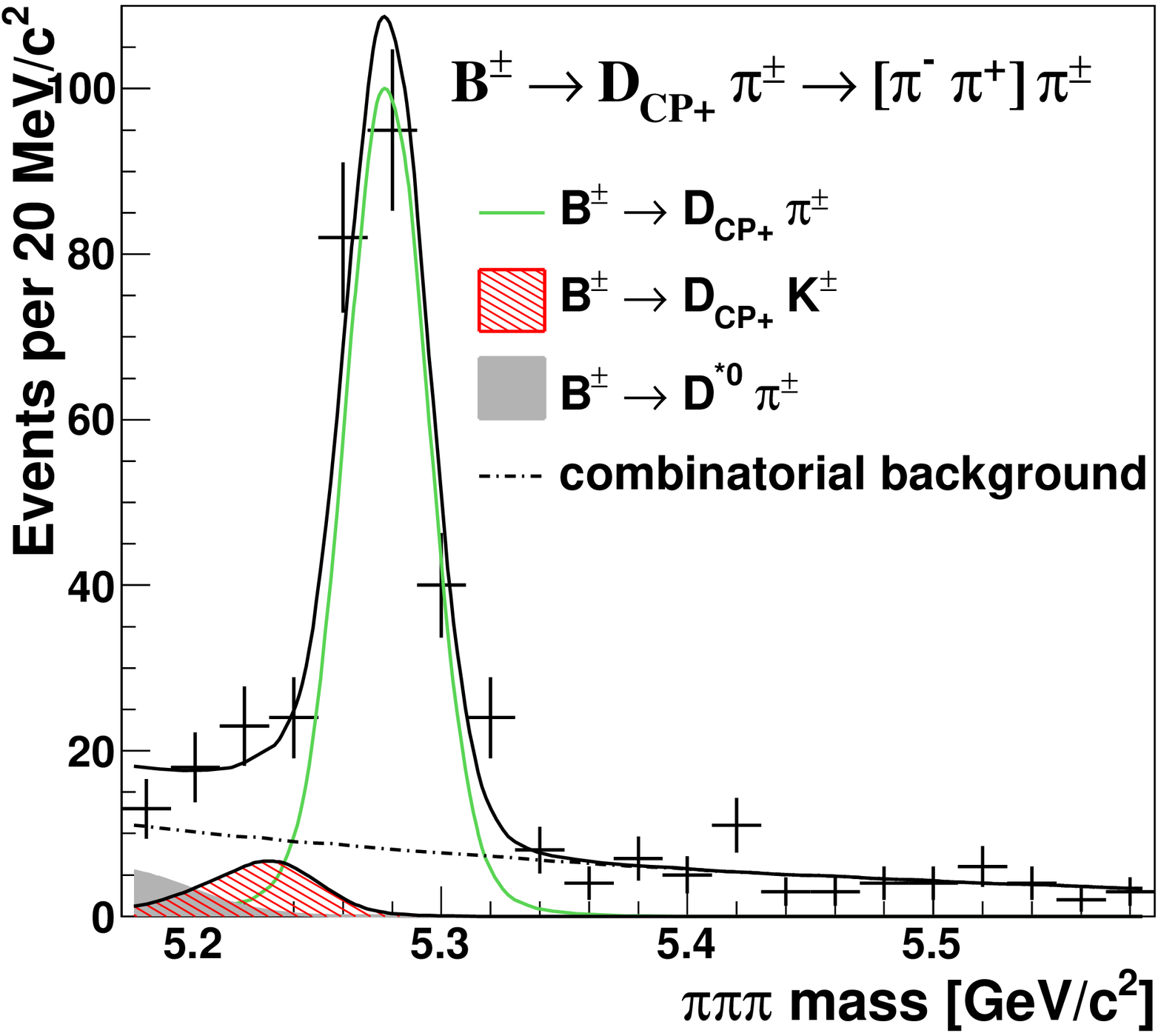}
\caption{Invariant mass distributions of $B^{-} \rightarrow D \pi^{-}$
candidates for each reconstructed decay mode. The pion mass is assigned to the charged track from the $B$ candidate decay vertex. The
projections of the common likelihood fit (see text) are overlaid for each mode.}
\label{mass}
\end{center}
\end{figure*}

For every $B^{-}\to D h^{-}$ candidate, a nominal invariant mass is
evaluated by assigning the charged pion mass to the particle $h^{-}$
coming from the $B$ decay.
The distributions obtained for the three modes of interest ($D\rightarrow K\pi, KK \textrm{ or } \pi\pi$) are reported in Fig.  \ref{mass}. A clear
$B^{-} \rightarrow D \pi^{-}$ signal is seen in each.
Events from $B^{-} \to D K^{-}$ decays are expected to form much smaller 
and wider peaks in these plots, located about 50 MeV/$c^{2}$ below 
the $B^{-} \rightarrow D \pi^{-}$ peaks, and as such cannot be resolved.
The dominant residual backgrounds are random track combinations that meet the selection requirements (combinatorial background), misreconstructed physics background such as $B^- \rightarrow D^{*0} \pi^-$ decay, and, in the $D^0\rightarrow KK$ final state, the nonresonant $B^-\rightarrow K^+ K^- K^-$ decay, as determined by a study performed on CDF simulation.

We used an unbinned likelihood fit, exploiting kinematic and particle
identification information from the measurement of
$dE/dx$ in a similar way to~\cite{bhh1}, to separate statistically the $B^{-} \to D K^{-}$ contributions
from the $B^{-} \to D \pi^{-}$ signals and from the combinatorial background.
To make best use of the available information, we fit the three modes 
simultaneously using a single likelihood function, to take advantage of the 
presence of parameters common to the three modes.

The likelihood function is

\begin{eqnarray}\label{eq:likelihood}
    \mathcal{L} & = & \prod_i (1-b)\sum_{j} f_j \mathcal{L}^{\mathrm{kin}}_j  \mathcal{L}^{\mathrm{PID}}_j + b\mathcal{L}^{\mathrm{kin}}_{\mathrm{c}}    \mathcal{L}^{\mathrm{PID}}_{\mathrm{c}}
\end{eqnarray}
where $c$ labels combinatorial background quantities, $b$ is the
combinatorial background fraction, and $\mathcal{L}^{\mathrm{kin}}$ and $\mathcal{L}^{\mathrm{PID}}$ are defined below. The index $j$ runs over the modes $B^{-} \rightarrow D
K^{-}$, $B^{-} \rightarrow D \pi^{-}$, nonresonant $B^- \rightarrow K^+ K^- K^{-}$ and $B^- \rightarrow \pi^+ \pi^- K^{-}$, and $B^{-} \rightarrow D^{*0} \pi^{-}$
(where a soft $\gamma$ or $\pi^0$ from the $D^{*0}$ is undetected) and $f_j$ are the fractions to be determined by the fit. The fraction of the physics background ($B^{-} \rightarrow D^{*0} \pi^{-}$) with respect to the signal is common to the three decays and the fraction of the $B^- \rightarrow D_{\mathit{CP+}} \pi^-$ is common to the two $D_{\mathit{CP}}$ modes.
As determined from simulation, these modes are the only significant contributions within the mass range
$5.17< M < 5.60$ GeV/$c^2$ chosen for our fit.

Kinematic information is given by three loosely correlated
observables: (a) the mass $M_{D \pi}$, calculated by assigning the pion
mass to the track from the $B$ decay; (b) the momentum imbalance 
$\alpha$,  defined as
  \begin{eqnarray}
  \alpha & = & 1-p_{tr}/p_{D} > 0 \qquad \mathrm{if}  \quad p_{tr}  <  p_{D};\nonumber\\
 \alpha &= & -(1-p_{D}/p_{tr}) \le 0 \qquad \mathrm{if}  \quad p_{tr}  \ge  p_{D};\nonumber
  \end{eqnarray}
  where $p_{tr}$ is the momentum of the track from the $B$ candidate; and (c) the scalar
  sum of the $D$ momentum and the momentum of the track from the $B$ candidate
  ($p_{tot} = p_{tr} + p_{D}$).  
  The above variables uniquely identify the invariant mass $M_{DK}$ evaluated with a 
  kaon mass assignment to the track from the $B$ decay, 
  through the (exact) relations~\cite{Squillacioti-Thesis}

 \begin{widetext}

{\small
\begin{eqnarray*}
{M^2_{DK} = M_{D\pi}^2 + m_{\pi}^2 - m_K^2 + 2
\sqrt{m_{D}^2 + \frac{p_{tot}^{2}}{(2-\alpha)^{2}}}
} \left(\sqrt{m_{\pi}^2 + \left(\frac{p_{tot}
(1-\alpha)}{2-\alpha}\right)^2} - \sqrt{m_K^2 + \left(\frac{p_{tot}
(1-\alpha)}{2-\alpha}\right)^2}\right)\nonumber
\end{eqnarray*}
}
if $\alpha > 0$;\\
{\small
\begin{eqnarray*}
{M^2_{DK} = M_{D\pi}^2 + m_{\pi}^2 - m_K^2 + 2
\sqrt{m_{D}^2 + \left(\frac{p_{tot}(1+\alpha)}{2+\alpha}\right)^2}}
 \left(\sqrt{(m_{\pi}^2 + \left(\frac{p_{tot}}{2+\alpha}
\right)^2} - \sqrt{m_K^2 +
\left(\frac{p_{tot}}{2+\alpha}\right)^2}\right)\nonumber
\end{eqnarray*}
}
if $\alpha \le 0$.\\
\end{widetext}
Using these variables, we can write $\mathcal{L}^{\mathrm{kin}}_j = P_j(M_{D\pi}|\alpha,p_{tot}) P_j(\alpha,p_{tot})$ and $\mathcal{L}^{\mathrm{PID}}_j = P_j(dE/dx|\alpha,p_{tot})$, where $P_j$ is the probability density function for decay mode $j$.
Distributions of the kinematic variables for the signals are obtained from samples of events from the full CDF simulation, while for the combinatorial background they are 
obtained from the mass sidebands of data. The shape of the mass distribution assigned to each signal process
($B^{-} \rightarrow D \pi^{-}$ and $B^{-} \rightarrow D K^{-}$ decays) has been
modeled in detail from a dedicated study including the effect of final state QED radiation~\cite{bhhPRD}. 
The simulation results were tested on high-statistics data samples of $D^0$ decays, in order
to ensure the reliability of the extraction of the $DK^{-}$ component 
in the vicinity of the larger $D\pi^{-}$ peak. Exponential functions were used to model the mass distribution of combinatorial background for each mode. The normalization and the slope of these functions are independently determined in the maximum likelihood fit. The particle identification (PID) model of the combinatorial background allows for pion and kaon components, which are free to vary in the fit. 

A large sample of $D^{*+}\to D^0(\to K^- \pi^+)\pi^+$ decays was used to calibrate the \dedx\ response of the detector 
to kaons and pions, using the charge of the pion in the $D^{*+}$ decay to determine the identity of the $D^0$ decay products. 
The calibration includes the dependence of the shape and the average of the response curve on
particle momentum, and the shape of the distribution of common-mode
fluctuations.  The calibrated \dedx\ information provides a 1.5 $\sigma$ separation power between pion and kaon particles of $p_T> 2$~GeV/$c$.
Uncertainties on the calibration parameters are included in the final systematic
uncertainty of $A_{\mathit{CP+}}$ and $R_{\mathit{CP+}}$~\cite{Squillacioti-Thesis}.

The $B^{-} \rightarrow D K^{-}$ and $B^{-}\rightarrow D\pi^{-}$ signal
event yields obtained from the fit to the data are
reported in Table \ref{fit_res}. The fraction of the $B^- \rightarrow \pi^+ \pi^- K^{-}$ was set by the fit to its lower bound at zero, compatible with the expectation of a negligible contribution, and will be ignored in the following.
The uncorrected values of the double ratio of branching
fractions $R_{\mathit{CP+}}$ and of the $\mathit{CP}$ asymmetry $A_{\mathit{CP+}}$ obtained from the
fit are $R_{\mathit{CP+}} = 1.27 \pm 0.24$ and $A_{\mathit{CP+}} = 0.39\pm 0.17$.
In the fit, $R_{\mathit{CP+}}$ and $A_{\mathit{CP+}}$ are functions of the fractions [$f_j$ in Eq. \ref{eq:likelihood}] and the total number of events in each subsample.
 
\begin{table*}[htb]
 \caption{$B^{-} \rightarrow D K^{-}$ and $B^{-}\rightarrow D \pi^{-}$
 event yields obtained from the fit to the data.}
  \begin{center}
    \begin{tabular}{c c c c c c c}
      \hline\hline
      \footnotesize{$D$ mode} & \footnotesize{$B^+\rightarrow D \pi^+$} &
      \footnotesize{$B^-\rightarrow D \pi^-$} & \footnotesize{$B^+\rightarrow D
      K^+$} & \footnotesize{$B^-\rightarrow D K^-$} & 
      \footnotesize{$B^+ \rightarrow [h^- h^+]K^+$} & 
      \footnotesize{$B^- \rightarrow [h^- h^+]K^-$}\\
      \hline
      $K^-\pi^+$ & $3769 \pm 68$ & $3763 \pm 68$ & $250 \pm 26$ & $266 \pm 27$ & - & -\\
      $K^+K^-$ & $381 \pm 25$ & $399 \pm 26$ & $22 \pm 8$ & $49 \pm 11$ & $3 \pm 1$ & $3 \pm 1$\\
      $\pi^+\pi^-$ & $101 \pm 13$ & $117 \pm 14$ & $6 \pm 6$ & $14 \pm 6$ & - & -\\
      \hline\hline
    \end{tabular}
    \label{fit_res}
  \end{center}
\end{table*}

\begin{figure*}[htbp]
\begin{center}
\includegraphics[width=2in]{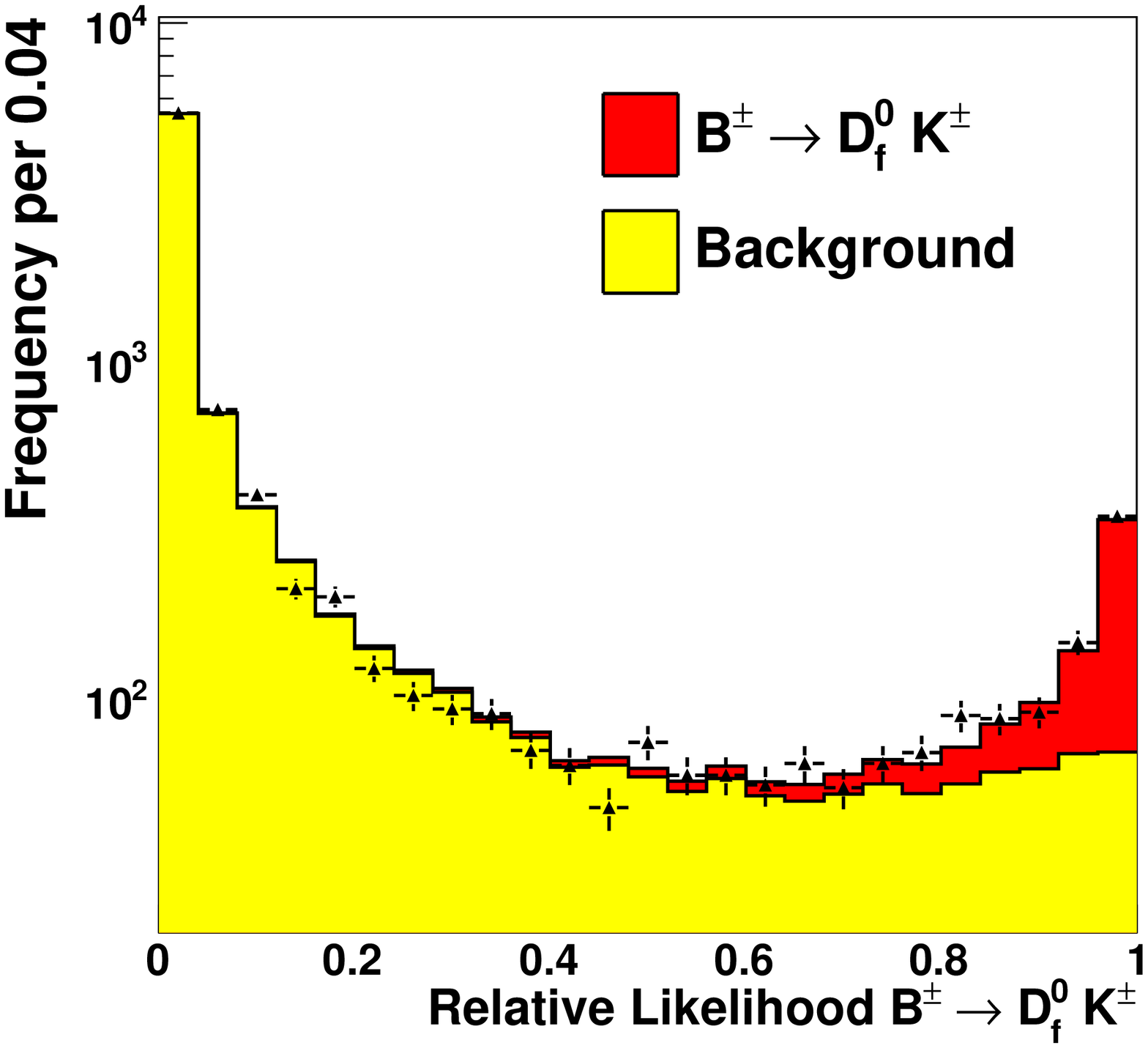}
\includegraphics[width=2in]{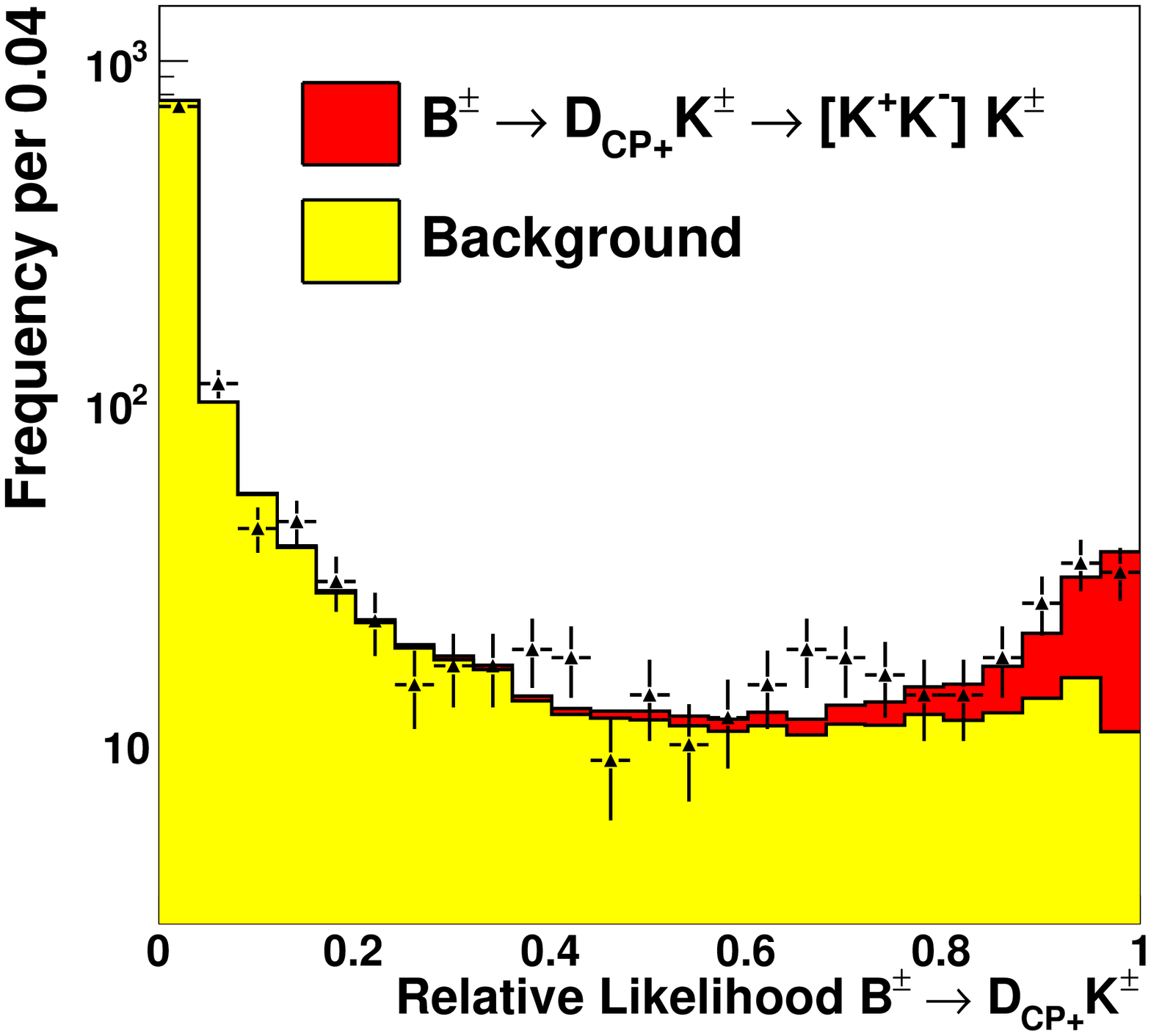}
\includegraphics[width=2in]{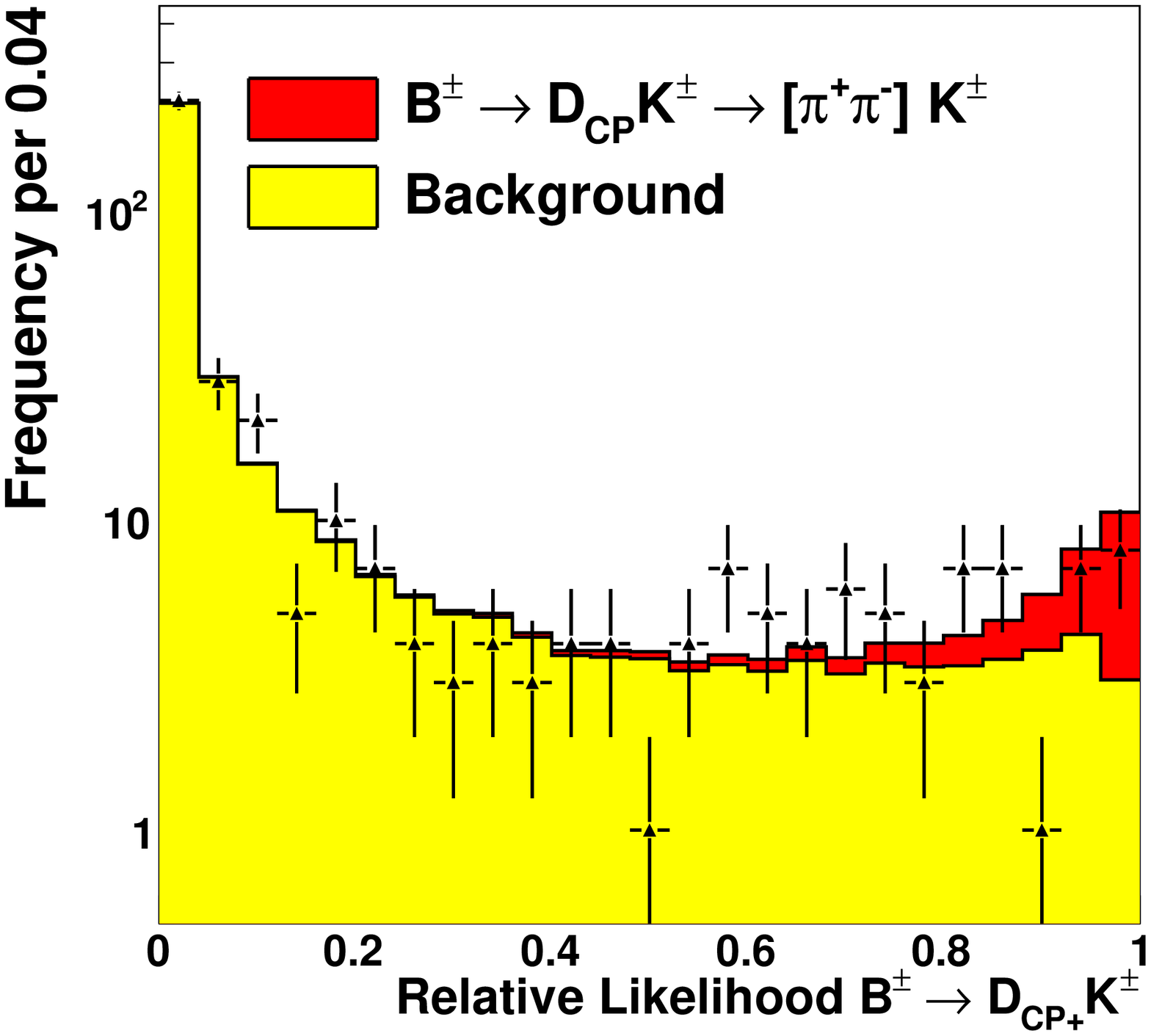}
\caption{Relative likelihood for $B^{-} \rightarrow D K^{-}$ candidates for each reconstructed decay mode. The points with the error bars show the distribution obtained on the fitted data sample while the histograms show the distributions obtained by generating signal and background events directly from the total PDF of the fit composition.}
\label{RL}
\end{center}
\end{figure*}

\begin{figure*}[htbp]
   \centering
   \includegraphics[width=2.4in]{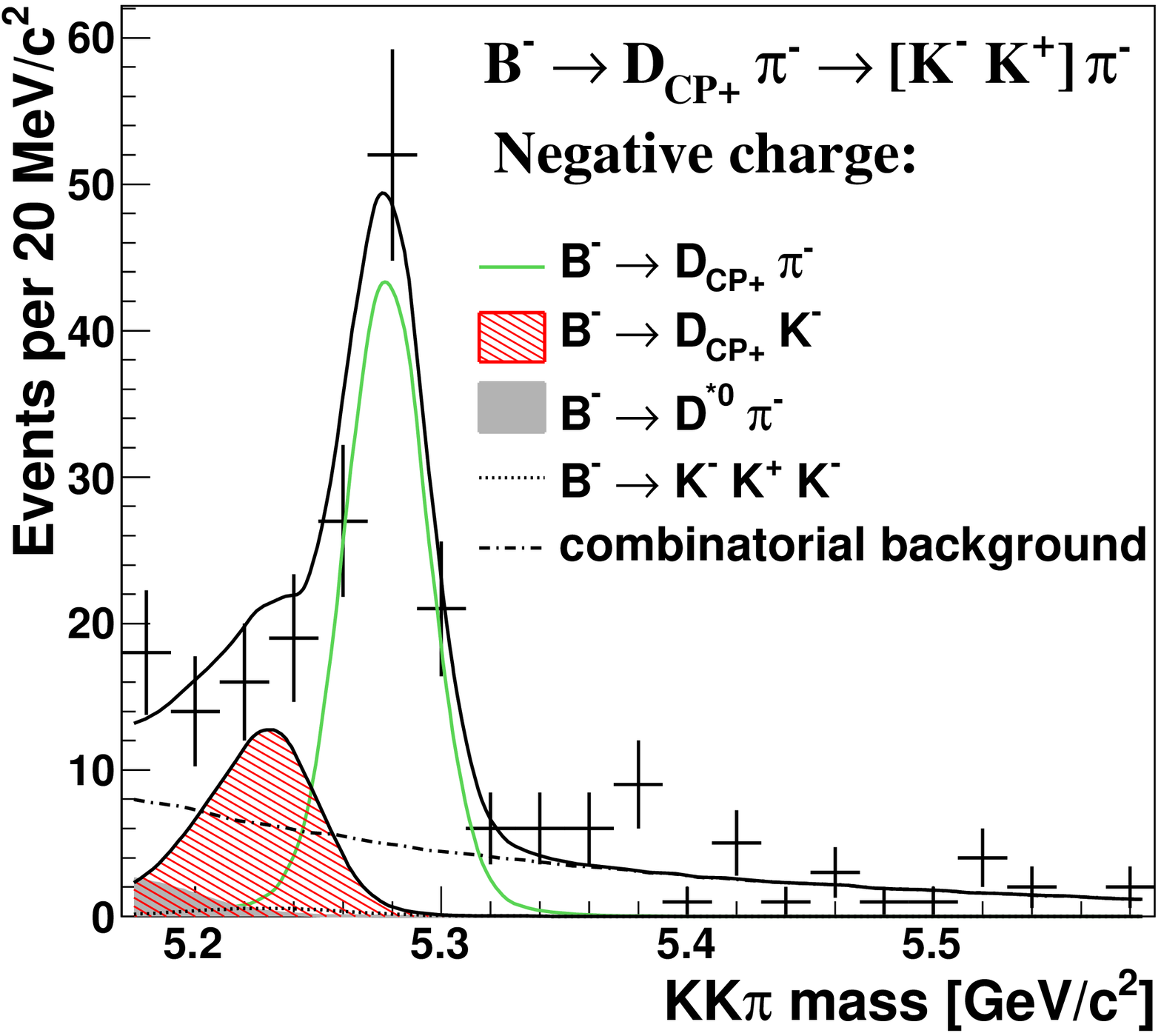}
   \includegraphics[width=2.4in]{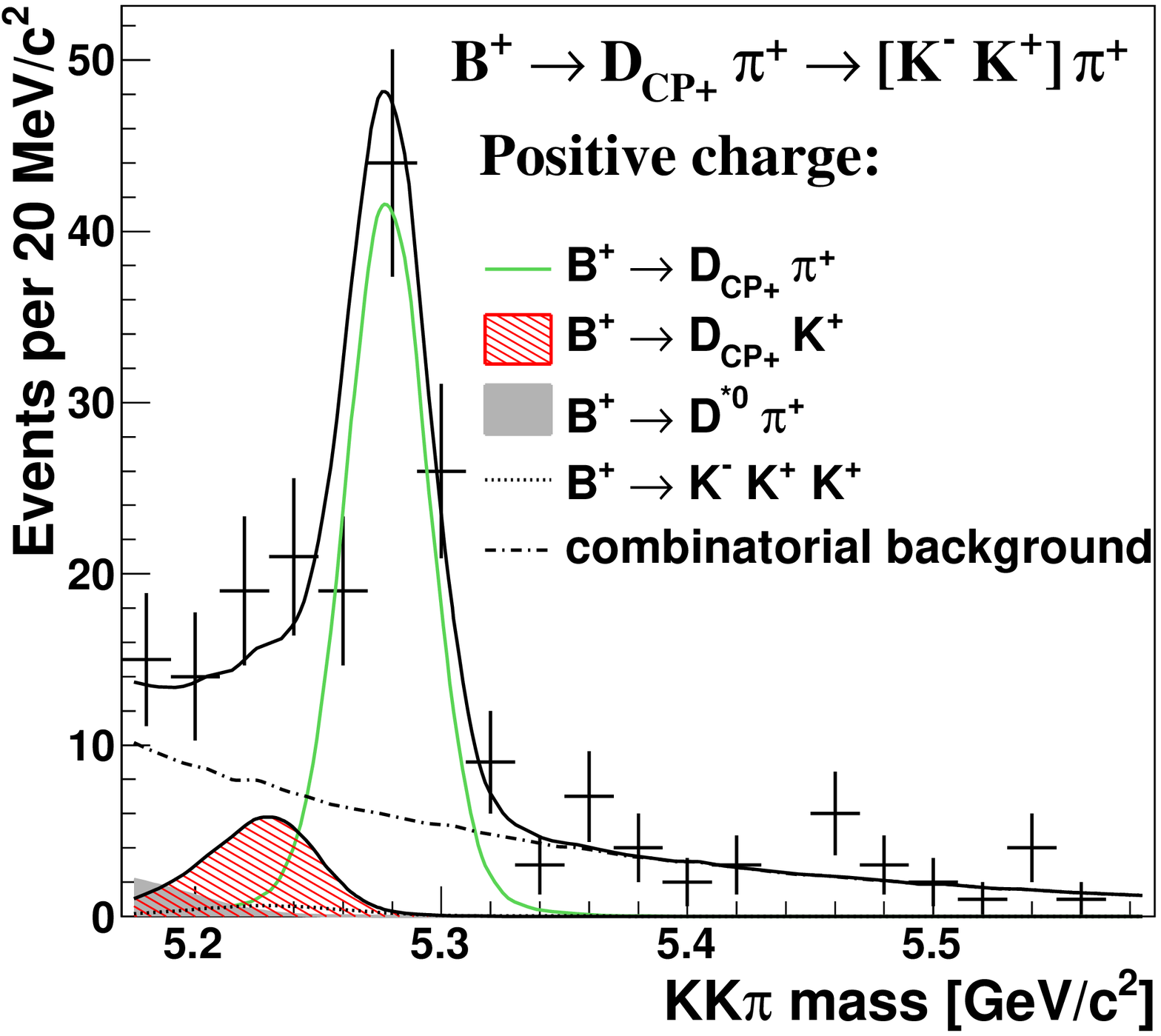}
   \includegraphics[width=2.4in]{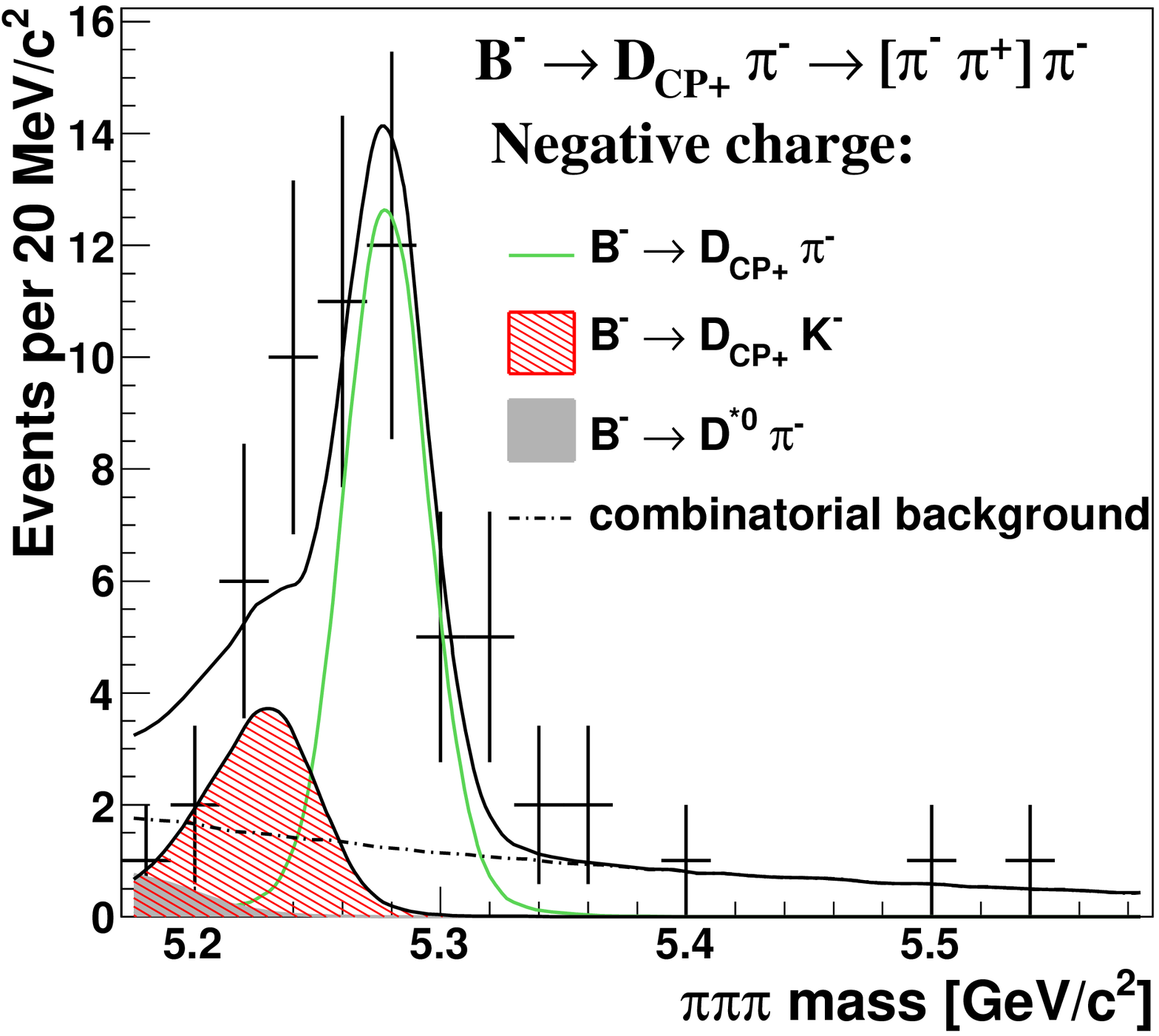}
   \includegraphics[width=2.4in]{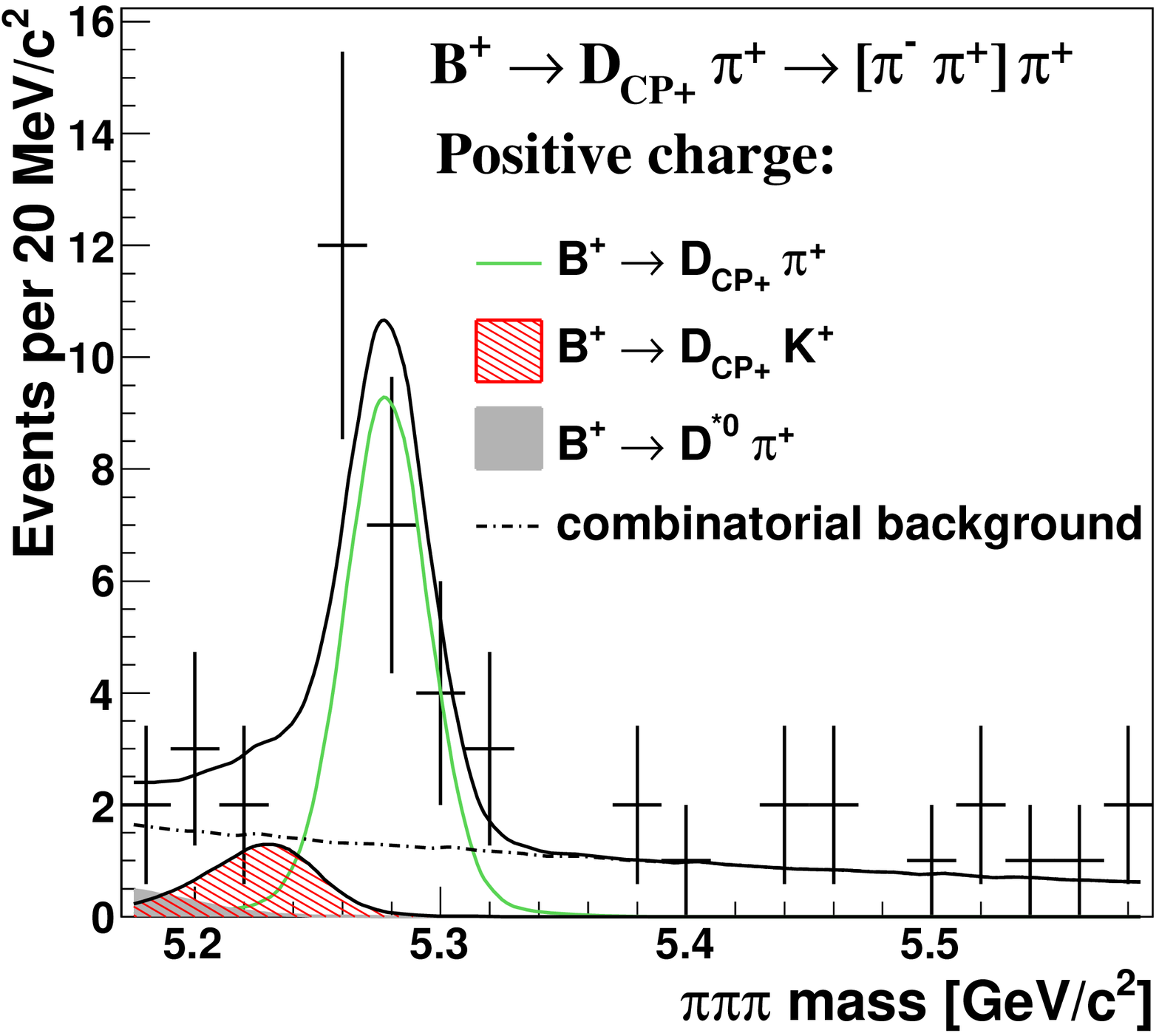}
    \caption{Invariant mass distributions of $B^{-} \rightarrow D
    \pi^{-}$ candidates for each reconstructed decay mode.  The pion
    mass is assigned to the prompt track from the $B$ decay.  A
    requirement on the PID variable was applied to suppress the $D\pi$
    component and favor the $DK$ component.  The projections of the
    likelihood fit for each mode are overlaid. The $p$--value for agreement of data with the fit is $0.95$}
   \label{proj_ID05}
\end{figure*}

As a check of the goodness of the fit, and to visualize better the separation between signal and
background, we plot distributions of the relative signal likelihoods:
\begin{equation}
RL=\frac{pdf(B\rightarrow DK)}{pdf(B\rightarrow DK)+pdf(background)}
\end{equation}
where $pdf(B\rightarrow DK)$ is the probability density
under the signal hypothesis, and $pdf(background)$ is the
probability density under the background
hypothesis (including both physics and combinatorial backgrounds, with their measured
relative fractions).  These distributions are compared to the 
prediction of our fit in Fig.~\ref{RL}, showing a very good agreement. 
In addition, we plot projections of the fit on the invariant mass 
distributions, both for the entire sample (Fig.~\ref{mass}), and for 
a kaon--enriched subsample, where the interesting $B^{-} \rightarrow D K^{-}$ 
components have been enhanced with respect to the $B^{-} \rightarrow D \pi^{-}$ by 
means of a \dedx\ cut (Fig.~\ref{proj_ID05}). All these projections show very good agreement between our fit and the data.

Some corrections are needed to convert our fit results into 
measurements of the parameters of interest. First, we correct for small biases in the fit 
procedure itself, as measured by repeated fits on simulated samples: 
$\delta(R_{\mathit{CP+}}) = -0.027\pm 0.005$ and $\delta(A_{\mathit{CP+}}) = 0.015\pm 0.003$. 
These biases are independent of the true values of $A_{\mathit{CP+}}$ and $R_{\mathit{CP+}}$ used 
in the simulated samples.
$R_{\mathit{CP+}}$ does not need any further corrections because detector effects cancel in the
double ratio of branching fractions. 
The direct $\mathit{CP}$ asymmetry $A_{\mathit{CP+}}$ needs to be corrected for 
the different probability for $K^+$ and $K^-$ mesons to interact with the
tracker material.  This effect is reproduced well by CDF II detector simulation (traced by GEANT
\cite{Brun}), which yields an estimate
$\frac{\epsilon(K^+)}{\epsilon(K^-)}=1.0178\pm 0.0023 (\mathrm{stat}) \pm 0.0045 (\mathrm{syst})$ 
\cite{acp_corr} which has been verified by measurements on data \cite{procMorello}.

The corrected results are

\begin{eqnarray}
R_{\mathit{CP+}} = 1.30\pm 0.24(\mathrm{stat}),\\
A_{\mathit{CP+}} = 0.39\pm 0.17(\mathrm{stat}),
\end{eqnarray}
where $A_{\mathit{CP+}}$ was corrected using the following equation:

{\small
\begin{equation}
A_{\mathit{CP+}}=\frac{N(B^-\rightarrow D^0_{\mathit{CP+}} K^-)\frac{\epsilon(K^+)}{\epsilon(K^-)}-N(B^+\rightarrow D^0_{\mathit{CP+}}K^+)}{N(B^-\rightarrow D^0_{\mathit{CP+}} K^-)\frac{\epsilon(K^+)}{\epsilon(K^-)}+N(B^+\rightarrow D^0_{\mathit{CP+}}K^+)}\nonumber.
\end{equation}}

Systematic uncertainties are listed in Table~\ref{systematic}. 
 They were determined by generating simulated samples of pseudoexperiments with different underlying assumptions, and checking the effect of such changes on the results of our measurement procedure. 
The dominant contributions are
uncertainty on the \dedx\ calibration and parametrization, uncertainty on the kinematics of the combinatorial background, and uncertainty on the
physics background ($B^{-}\rightarrow D^{*0}\pi^{-}$) mass distribution. 
Variations in the model of the combinatorial background included different functional forms of the mass distribution, and alternative ($\alpha$, $p_{tot}$) distributions, constrained by comparison with real data in the mass sidebands.

Smaller
contributions are assigned for trigger efficiencies, assumed $B^-$ mass input in the fit \cite{bumass} and kinematic properties of signal and physics background. 

\begin{table}[htb]
 \caption{Summary of systematic uncertainties.}
  \begin{center}
    \begin{tabular}{c c c}
       \hline\hline
      Source & $R_{\mathit{CP+}}$ & $A_{\mathit{CP+}}$\\
      \hline
      \dedx\ model & 0.056 & 0.030\\
      $D^{*0}\pi$ mass model & 0.025 & 0.006\\
      Input $B^-$ mass to the fit & 0.004 & 0.002\\
      Combinatorial background mass model & 0.020 & 0.001\\
      Combinatorial background kinematics & 0.100 & 0.020\\
      $D \pi$ kinematics & 0.002 & 0.001\\
      $D K$ kinematics & 0.002 & 0.004\\
      $D^{*0} \pi$ kinematics & 0.004 & 0.002\\
      Fit bias & 0.005 & 0.003\\
       \hline
      Total (sum in quadrature) & 0.12 & 0.04\\
      \hline\hline\\
    \end{tabular}
    \label{systematic}
  \end{center}
\end{table}

In summary,
we have measured the double ratio of $\mathit{CP}$-even to flavor eigenstate branching fractions [Eq.  \ref{rcp_eq}]
$R_{\mathit{CP+}} = 1.30\pm 0.24(\rm{stat})\pm 0.12(\rm{syst})$ and the direct $\mathit{CP}$ asymmetry [Eq. \ref{acp_eq}] $A_{\mathit{CP+}} = 0.39\pm 0.17(\rm{stat})\pm 0.04(\rm{syst})$.
These results can be combined with other $B^{-} 
\rightarrow D K^{-}$ decay parameters to improve the  
determination of the CKM angle $\gamma$. These measurements are performed here for the first time in 
hadron collisions, are in agreement with previous measurements from 
$BaBar$ ($R_{\mathit{CP+}} = 1.06 \pm 0.10 \pm 0.05$, $A_{\mathit{CP+}} 
= 0.27 \pm 0.09 \pm 0.04$ in 348~fb$^{-1}$ of integrated luminosity \cite{babar}) and Belle ($R_{\mathit{CP+}} = 1.13 \pm 0.16 \pm 0.08$, $A_{\mathit{CP+}} = 0.06 \pm 0.14 \pm 0.05$ in 250~fb$^{-1}$ of integrated luminosity \cite{belle})
and have comparable uncertainties.

 
\section*{Acknowledgments}
We thank the Fermilab staff and the technical staffs of the participating institutions for their vital contributions. This work was supported by the U.S. Department of Energy and National Science Foundation; the Italian Istituto Nazionale di Fisica Nucleare; the Ministry of Education, Culture, Sports, Science and Technology of Japan; the Natural Sciences and Engineering Research Council of Canada; the National Science Council of the Republic of China; the Swiss National Science Foundation; the A.P. Sloan Foundation; the Bundesministerium f\"ur Bildung und Forschung, Germany; the Korean Science and Engineering Foundation and the Korean Research Foundation; the Science and Technology Facilities Council and the Royal Society, UK; the Institut National de Physique Nucleaire et Physique des Particules/CNRS; the Russian Foundation for Basic Research; the Ministerio de Ciencia e Innovaci\'{o}n, and Programa Consolider-Ingenio 2010, Spain; the Slovak R\&D Agency; and the Academy of Finland.

 \bibliography{bibliografia}

\begin{thebibliography}{31}
\bibitem[{decay()}]{chargedecay}
  {\bibinfo{title}{{The charge conjugate state is implied throughout the paper, except in formulas and sentences where both are mentioned explicitly.}}}
\bibitem[{CDF()}]{CDF-coordinates}
  {\bibinfo{title}{{CDF II uses a cylindrical coordinate system in which
        $\phi$ is the azimuthal angle, $r$ is the radius from the nominal beam line,
        and $z$ points in the proton beam direction, with the origin at the center of
        the detector. The transverse plane is the plane perpendicular to the $z$
        axis.}}}
\bibitem[{Iso()}]{Isolation}
  {\bibinfo{title}{{Isolation is defined as
        $I_{B}=p_T(B)/(p_T(B)+\sum_ip_{Ti})$, where $p_T(B)$ is the transverse
        momentum of the $B$ candidate, and the sum runs over all other tracks within
        a cone of radius 1, in $\eta-\phi$ space around the $B$ flight-direction. 
        Its value is typically higher for bottom--flavored hadrons than for random track combinations.}}}
\end{thebibliography}
 
 \end{document}